\documentclass[12pt,preprintnumbers,eqsecnum]{revtex4}
\usepackage{amssymb}
\usepackage{amsthm}
\usepackage{pstricks}
\usepackage{color}
\usepackage{slashed}
\usepackage{newlfont}
\usepackage{ulem}
\usepackage{array}

\usepackage{float}
\usepackage{amsmath,amsfonts,graphicx,bbm,bm}
\usepackage{subfigure}

\def\beq{\begin{equation}}
\def\eeq{\end{equation}}
\def\sp{\frac{3}{2}}
\def\sh{\hat{s}}

\def\bt{\begin{table}}
\def\et{\end{table}}
\def\bc{\begin{center}}
\def\ec{\end{center}}
\def\bi{\begin{itemize}}
\def\ei{\end{itemize}}
\def\bea{\begin{eqnarray}}
\def\eea{\end{eqnarray}}
\def\beas{\begin{eqnarray*}}
\def\eeas{\end{eqnarray*}}

\begin{document}

\setlength{\parskip}{0.1cm}

\preprint{OSU-HEP-12-10}

\title{\large {\sc Search for spin--3/2 quarks at the 
                  Large Hadron Collider} }

\author{Duane A. Dicus$^{1,}$\footnote{Electronic address: dicus@physics.utexas.edu},
Durmus Karabacak$^{2,}$\footnote{Electronic address: durmas@ostatemail.okstate.edu},
S. Nandi$^{2,}$\footnote{Electronic address: s.nandi@okstate.edu}, and
Santosh Kumar Rai$^{2,}$\footnote{Electronic address: santosh.rai@okstate.edu}}
\affiliation{$^1$Department of Physics, \\ 
   University of Texas, Austin TX 78712-1081, USA \\
$^2$Department of Physics and Oklahoma Center for High Energy Physics, \\
Oklahoma State University, Stillwater OK 74078-3072, USA}

  
 

\begin{abstract}
We consider the pair production of color triplet spin--$\frac{3}{2}$ 
quarks and their subsequent decays at the LHC. This particle, if 
produced, will most likely decay into  top quark and gluon, bottom 
quark and gluon, or a light quark jet and gluon, depending on the 
quantum number of the spin--$\frac{3}{2}$ particle. This would lead to 
signals with $t\bar{t}jj$, $b\bar{b}jj$, or $4j$ in the 
final states. We present a detailed analysis of the signals 
and backgrounds at $\sqrt{s}= 7$, $8$ and $14$ TeV and show 
the reach for such particles by solving for observable mass values for the 
spin--$\frac{3}{2}$ quarks through its decay products. 
\end{abstract}

\maketitle

\section{\label{sec:intro}Introduction}

The Standard Model (SM) of particle physics has been extensively tested 
by many independent experiments and the results are in 
agreement with the predictions of the SM. The Large Hadron 
Collider (LHC) at CERN is designed to explore the energy and intensity frontier which
could show physics beyond the SM. The initial 
results released by the ATLAS and CMS experiments not only confirm 
the predictions of the SM, including the discovery of the Higgs boson \cite{cms:2012gu,atlas:2012gk}, 
but have also started pushing the energy scale required by new physics models including
exotic fermions and gauge bosons 
which are not present in the SM. Among exotic fermions one possible new 
particle is a spin--$\frac{3}{2}$  excitation of quarks. We 
will assume this spin--$\frac{3}{2}$ particle to be a color triplet 
like an ordinary quark and consider the pair production and the decay 
of such an exotic particle at the LHC. 
 
It is not outside the realm of possibility that a  spin--$\frac{3}{2}$ 
quark could exist as a fundamental particle.  We could also have 
spin--$\frac{3}{2}$ bound states of ordinary quarks with gluons or the 
Higgs boson.  There are also theoretical models in which spin--$\frac{3}{2}$ 
quarks arise as bound states of three heavy quarks for sufficiently 
strong Yukawa couplings \cite{tay}.  The masses of these bound states 
are typically expected to be a few TeV. A heavy spin--$\frac{3}{2}$  
quark could also exist as the lightest Regge recurrences of light 
spin--$\frac{1}{2}$ quarks or as Kaluza-Klein modes 
in string theory if one or more of the compactification radii is of the 
order of the weak scale rather than the Planck scale and such weak 
compactification in the framework of both string theory and field theory 
has been popular \cite{sure}. In this work we restrict ourselves 
to the collider production of point-like spin--$\frac{3}{2}$ color triplet 
quarks. The production of spin--$\frac{3}{2}$ quarks by hadronic collisions has been 
previously considered by Moussallam and Soni \cite{mous} and by Dicus, Gibbons, and Nandi
\cite{Dicus:1998yc}. There are several studies 
on production of spin--$\sp$ fermions at lepton 
colliders \cite{Walsh:1999pb,Almeida:1995yp,Cakir:2007wn} and also the virtual
effects of such particles on $t\bar{t}$ production \cite{Stirling:2011ya}.

Our paper is organized as follows.  In Section \ref{sec:frules}, we give  
the Feynman rules relevant for the production of spin--$\frac{3}{2}$ quarks. 
In Section \ref{sec:cross}, we give the explicit analytic formulae 
for the squares of the amplitude, various subprocess cross sections and 
total production cross sections. In Section \ref{sec:signals}, 
we present the 
analysis of the signal of spin--$\frac{3}{2}$ particle decaying into light 
jets or into heavy flavor modes. Here we make the physics analysis 
of relevant background and signal for three different decay scenarios. 
Section \ref{sec:summary} contains a summary.  

\section{\label{sec:frules}Feynman Rules for Spin--$\frac{3}{2}$ Particles}

The Lagrangian and the equations of motion for a free 
spin--$\sp$ particle of mass $M$ can be written as \cite{rari, mold} 
\begin{equation}\label{eq:lag}
\mathcal{L} = \bar{\psi}_{\alpha} \Lambda^{\alpha\beta}\psi_\beta
\end{equation} 
\begin{equation}
\Lambda^{\alpha\beta}\psi_\beta=0
\end{equation}
where
\begin{equation}
\Lambda_{\alpha\beta} =(i\slashed \partial- M)g_{\alpha\beta}
          + iA(\gamma_\alpha \partial_\beta+\gamma_\beta \partial_\alpha)
         +\frac{iB}{2}\gamma_\alpha \slashed \partial \gamma_\beta 
         + CM\gamma_\alpha \gamma_\beta
\end{equation}
with $B \equiv 3A^2+2A+1$ and $C \equiv 3A^2+3A+1$. The parameter $A$ is 
arbitrary except that $A \not=-\frac{1}{2}$. The field $\psi_\alpha$ 
satisfies the subsidiary conditions
\begin{align}
\gamma^\alpha \psi_\alpha &= 0  \label{R1}\\
\partial^\alpha \psi_\alpha & =0. \label{R2}
\end{align}

The 
propagator $S_{\alpha\beta}$ is given by
\begin{equation}
\begin{split}
S_{\alpha\beta}(p)=&\frac{1}{\slashed p -M} \bigg[g_{\alpha\beta}
            - \frac{1}{3}\gamma_\alpha \gamma_\beta 
            - \frac{2}{3 M^2}p_\alpha p_\beta 
            + \frac{1}{3M}(p_\alpha \gamma_\beta
            - p_\beta \gamma_\alpha)\bigg]\\
    & + \Bigg\{\frac{a^2}{6M^2} \slashed p \gamma_\alpha \gamma_\beta 
      - \frac{ab}{3M}\gamma_\alpha \gamma_\beta 
      + \frac{a}{3M^2}\gamma_\alpha p_\beta 
      + \frac{ab}{3M^2}\gamma_\beta p_\alpha \Bigg\}
\end{split}
\end{equation}
where
\begin{eqnarray*}
a =\frac{A+1}{2A+1} \quad \mbox{and} \quad
b =\frac{A}{2A+1} ~~ .
\end{eqnarray*}
From Eq.(\ref{R1}) and Eq.(\ref{R2})
the terms depending on the parameter $A$ in the 
propagator vanish on the mass shell. 
A redefinition of the spin--$\frac{3}{2}$ field $\psi_\alpha$ allows one 
to remove the $A$ dependent terms in the propagator \cite{pasc}. However, 
in our analysis we have kept the $A$ dependence in the propagator and 
in the interaction vertices and used the disappearance of A as a check on our calculations.

The minimal substitution in 
Eq.(\ref{eq:lag}) gives the interaction of 
spin--$\frac{3}{2}$ quarks with gluon and photon fields, 
\begin{equation}
\mathcal{L}_I = g \bar{\psi}_\alpha \bigg( \frac{B}{2} \gamma^\alpha 
                  \gamma^\mu \gamma^\beta + A g^{\alpha\mu}\gamma^\beta
              + A \gamma^\alpha g^{\mu\beta} 
              + g^{\beta\alpha}\gamma^\mu \bigg) T_a \psi_\beta A_\mu^a\,\,,
\end{equation}
where $g$ is the coupling constant, $T_a$'s are the group generators and 
$A_\mu^a$ are the gauge fields.  For on-shell particles only the last term is nonzero.

\section{\label{sec:cross}Calculation of Cross Sections}
\begin{figure}[t!]
\centering
\includegraphics[width=6.6in,height=1.1in]{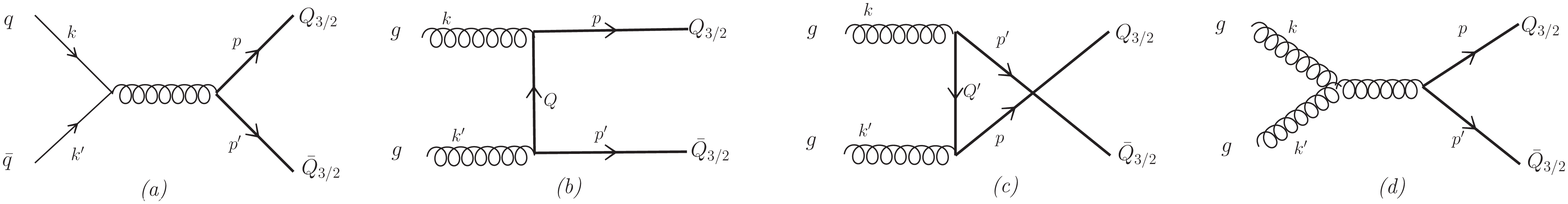}
\caption{\textit{The leading order (LO) Feynman diagrams for the pair production of 
spin--$\sp$ quarks through (a) $q\bar{q}$ and $gg$ initial states in (b) $t$-channel, 
(c) $u$-channel and (d) $s$-channel.}}
\label{feyndiag}
\end{figure}

In this section we 
provide the expressions necessary for the process,
\begin{equation}\label{eq:ppQQ}
p p \rightarrow Q_{3/2} \bar{Q}_{3/2} + X\,\,
\end{equation}
where $Q_{3/2}$ is the spin--$\sp$ quark. 
There are two subprocesses which contribute, 
$q\bar{q}$ annihilation and gluon fusion. 
The Feynman diagrams are shown in Fig.\ref{feyndiag} where \textit{(a)} represents
the $q\bar{q}$ annihilation while \textit{(b)--(d)} represent the $t$,$u$ and $s$-channel 
contributions of the gluon fusion subprocess respectively.
Just as for top quark production the largest 
contribution to the production of spin--$\sp$ at LHC energies is through gluon fusion. 

The t-channel amplitude shown in Fig.\ref{feyndiag} is given by 
\begin{equation}\label{eq:mt}
\begin{split}
\mathcal{M}_t =~ & g_s^2\bar{u}_\rho (p)\big(g^{\rho\alpha} \gamma^\mu  
        + A g^{\mu \rho} \gamma^\alpha \big) T_a \epsilon^{a}_\mu(k) \\
    &  \Bigg\{\frac{1}{\slashed Q -M}\Bigg[g_{\alpha\beta}
        - \frac{1}{3}\gamma_\alpha \gamma_\beta
        - \frac{2}{3 M^2}Q_\alpha Q_\beta 
        + \frac{1}{3M}\big(Q_\alpha \gamma_\beta 
        - Q_\beta \gamma_\alpha \big)\Bigg]\\
    &   + \frac{a^2}{6 M^2}\slashed Q \gamma_\alpha \gamma_\beta 
        - \frac{ab}{3 M}\gamma_\alpha \gamma_\beta 
        + \frac{a}{3M^2}\gamma_\alpha Q_\beta 
        + \frac{ab}{3M^2}\gamma_\beta Q_\alpha \Bigg\}\\
    &  \big(g^{\sigma\beta} \gamma^\nu + A g^{\nu\sigma} 
       \gamma^\beta \big) T_b \epsilon_\nu^b(k') v_\sigma(p')\quad ,
\end{split}
\end{equation}
while the $u$ channel amplitude has a similar form due to 
crossing symmetry,
\begin{equation}
\begin{split}
\mathcal{M}_u =~ & g_s^2\bar{u}_\rho (p)\big(g^{\rho\beta} \gamma^\nu  
          + A g^{\rho \nu} \gamma^\beta \big) T_b \epsilon^{b}_\nu(k') \\
    &  \Bigg\{\frac{1}{\slashed Q^{'} -M}\Bigg[g_{\beta\alpha}
       - \frac{1}{3}\gamma_\beta \gamma_\alpha
       - \frac{2}{3 M^2}Q^{'}_\beta Q^{'}_\alpha 
       + \frac{1}{3M}\big(Q^{'}_\beta \gamma_\alpha 
       - Q^{'}_\alpha \gamma_\beta \big)\Bigg]\\
    &  + \frac{a^2}{6 M^2}\slashed Q^{'} \gamma_\beta \gamma_\alpha 
       - \frac{ab}{3 M}\gamma_\beta \gamma_\alpha 
       + \frac{a}{3M^2}\gamma_\beta Q^{'}_\alpha 
       + \frac{ab}{3M^2}\gamma_\alpha Q^{'}_\beta \Bigg\}\\
    &  \big(g^{\sigma\alpha} \gamma^\mu + A g^{\sigma\mu} 
       \gamma^\alpha \big) T_a \epsilon_\mu^a(k) v_\sigma(p')\,,
\end{split}
\end{equation}
where $\slashed{Q}=\slashed{p}-\slashed{k}$, 
$\slashed{Q}'=\slashed{k}-\slashed{p}'$.
The amplitude for the $s$-channel contribution has a much simpler form 
because the $A$ dependence goes away for the spin--$\sp$ particles 
produced on-shell,
\begin{equation}\label{eq:ms}
\begin{split}
\mathcal{M}_s =& -ig_s^2 f_{abc} \bar{u}^\rho(p)\gamma^\alpha 
                 T^c v_\rho(p')\frac{1}{\hat{s}} \epsilon_{\mu}^a(k) \epsilon_{\nu}^b (k') \\
     & \bigg[g^{\mu\alpha}(2k+k^{'})^{\nu}
           - g^{\alpha\nu}(2k^{'}+k)^{\mu}
           + g^{\nu\mu}(k^{'}-k)^\alpha\bigg]~.
\end{split}
\end{equation}
The $\epsilon^a$'s represent the gluon fields 
while the spin--$\sp$ particles are denoted by the $u$ and $v$ spinors 
carrying Lorentz indices. From the expressions $\mathcal{M}_t$ and $\mathcal{M}_u$ we see that 
off-shell spin--$\sp$ particle exchange 
leads to an explicit dependence on the contact parameter $A$. Although 
we have this dependence in the amplitudes, the final results should be 
independent of $A$.  Indeed, we find that this dependence goes away not only from the 
final total result but also from each individual contribution such as 
$\Sigma |\mathcal{M}_t|^2$ or $\Sigma |\mathcal{M}_u|^2$ or the cross terms. 
This was verified by calculating the amplitude squares and all 
interference terms in both axial-gauge and Feynman-gauge. 

Using  Eqs.(\ref{eq:mt}-\ref{eq:ms}), the full spin and 
color averaged matrix amplitude square for the gluon-gluon subprocess 
is 
\begin{eqnarray} 
\sum |\mathcal{M}|^2_{GG} &=& \frac{g_s^4}{1944}\left[-2106
          - \frac{5832 M^2}{\sh}+\frac{112 \sh}{M^2}-\frac{272 \sh^2}{M^4}
          + \frac{39 \sh^3}{M^6}-\frac{2592 M^4 \sh^2}{u'^2 t'^2}
          - \frac{48 \sh^4}{u'^2t'^2} \right. \nonumber \\ 
&+& \left. 
            \frac{5832 M^4}{u' t'}+\frac{2592 M^2 \sh}{u' t'}
          + \frac{539 \sh^2}{u' t'}+\frac{4 \sh^3}{M^2 u' t'}
          + \frac{33 \sh^4}{M^4 u' t'}+\frac{521 u' t'}{M^4}
          + \frac{2916 u' t'}{\sh^2} 
\right. \nonumber \\ 
&-& \left. 
            \frac{121 \sh u' t'}{M^6}+\frac{4 \sh^2 u' t'}{M^8}
          - \frac{8 u'^2 t'^2}{M^8}\right]
\label{eq:matggsq}
\end{eqnarray}
where $t'$ and $u'$ are related to the usual definitions of the Mandelstam 
variables $t$ and $u$ in the parton center-of-mass frame as $t'=t-M^2$ and 
$u'=u-M^2$. The total cross section for the gluon-gluon subprocess is then 
\begin{equation}\label{eq:gg2QQbar}
\begin{split}
\hat{\sigma}(gg\rightarrow Q_{3/2} \bar{Q}_{3/2})
   =&\frac{\pi\alpha_s^2}{116640~ \hat{s}}
     \Bigg\{ 60 ~\ln\bigg(\frac{1+\beta}{1-\beta}\bigg) \bigg[66 y^2+8y+886
          + 5184\frac{1}{y} +1296\frac{1}{y^2}\bigg]
          \\          
    &+\beta\bigg[24 y^4+1178y^3-13626 y^2+11380 y -97200
                            -602640\frac{1}{y}\bigg]\Bigg\}
\end{split}
\end{equation}
where $\alpha_s \equiv g^2_s/4\pi$ , $y\equiv \hat{s}/M^2$ and 
$\beta \equiv \sqrt{1-4/y}$.
This expression for the total subprocess cross section  agrees with Ref.\cite{Dicus:1998yc}, but disagrees  with 
Ref.\cite{mous}. 
However Ref.\cite{mous} has an algebraic error which, when corrected, gives agreement with Eq.(\ref{eq:gg2QQbar}) \cite{mous2}.

The pair production of the spin--$\sp$ quarks will also have contributions 
coming from the amplitude for the quark-antiquark annihilation subprocess 
which is given by
\begin{equation}\label{eq:mqq}
\mathcal{M}_{q\bar{q}}=-i g^2\frac{1}{\hat{s}}\bar{u}^{\rho}(p)T_a\gamma^\mu v_\rho(p^{'})
                                         \bar{u}(k)\gamma_\mu v(k')~.
\end{equation}
The spin and color averaged matrix amplitude square for the quark-antiquark 
process is
\begin{eqnarray} 
\sum |\mathcal{M}_{q\bar{q}}|^2 &=& \frac{4g_s^4}{81 M^4 \sh^2} 
               \left[36 \sh M^6 - 2 \sh M^2  (\sh+2t')^2
             + \sh^2 (\sh^2+2 \sh t'+2 t'^2) \right. \nonumber \\ 
    &&\left. + 2 M^4 (\sh^2+18 \sh t'+18 t'^2)\right]
\label{eq:matqqsq}
\end{eqnarray}
and the total cross section for this subprocess is
\begin{equation}\label{eq:qq2QQbar}
\hat{\sigma}(q\bar{q}\rightarrow Q_{3/2} \bar{Q}_{3/2})=
             \frac{\pi \alpha_s^2}{81 \hat{s}}\beta 
       \Bigg[\frac{8}{3}y^2-\frac{16}{3}y-\frac{16}{3}+96\frac{1}{y}\Bigg]~.
\end{equation}

\begin{figure}[t!]
\centering
\includegraphics[width=3.4in]{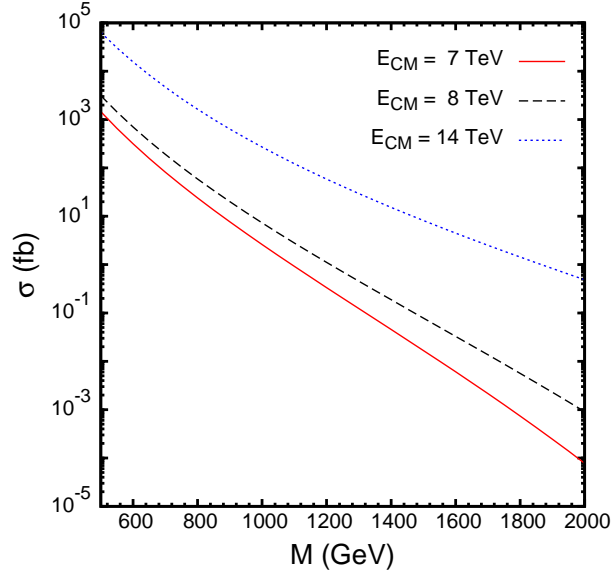}
\caption{\it The production cross sections for 
$p p \rightarrow Q_{3/2} \bar{Q}_{3/2}+X$ at the LHC as a function of 
spin--$\sp$ quark mass $M$ at center-of-mass energies, 
$E_{CM} = 7, 8$ and $14$ TeV. We have chosen the scale as 
$Q=M$, the mass of the spin--$\sp$ quark.}
\label{prodfig}
\end{figure}

To obtain the production cross section we convolute 
Eq.(\ref{eq:gg2QQbar}) and Eq.(\ref{eq:qq2QQbar}) with the parton 
distribution functions (PDF).
\begin{equation}\label{prodcros}
\begin{split}
\sigma({pp \rightarrow Q_{3/2} \bar{Q}_{3/2}+X})=& 
            \left\{ \sum_{i=1}^{5}  \int dx_1 \int dx_2 
        ~\mathcal{F}_{q_i}(x_1,Q^2)\times \mathcal{F}_{\bar{q}_i}(x_2,Q^2) 
         \times \hat{\sigma}(q_i\bar{q}_i\rightarrow Q \bar{Q}) \right\} \\
  & + \int dx_1 \int dx_2 ~\mathcal{F}_g (x_1,Q^2) \times 
                           \mathcal{F}_g (x_2,Q^2) 
                  \times \hat{\sigma}(gg\rightarrow Q \bar{Q}), 
\end{split}
\end{equation}
where $\mathcal{F}_{q_i}$, $\mathcal{F}_{\bar{q}_i}$ and $\mathcal{F}_{g}$  
represent the respective PDF's for partons (quark, antiquark and gluons) in 
the colliding protons, while $Q$ is the factorization scale.  
In Fig.\ref{prodfig} we plot the leading-order production cross section 
for the process $p p \rightarrow Q_{3/2} \bar{Q}_{3/2}+X$  at center of 
mass energies of 7, 8 and 14 TeV  as a function of the spin--$\sp$ quark 
mass $M$.
We set the factorization scale $Q$ equal to $M$,  
and used the {\tt CTEQ6$\ell$1} parton distribution 
functions \cite{Pumplin:2002vw}. This production 
cross section is larger than any spin-$\frac{1}{2}$ colored fermion of 
same mass such as a fourth-generation quark or an excited quark. This is 
not unexpected, as the cross section given in Eq.(\ref{eq:gg2QQbar}) 
grows with energy as $\hat{s}^3$ which violates 
unitarity at high energies. We assume that the interactions given in 
Sec.II represent an effective interaction such that, at higher energies, 
higher order contributions will be important and the cross section will 
be damped by some form factors dependent on the scale of the new physics. 
Some explicit ways to address this have been discussed in 
Refs.\cite{Hassanain:2009at, Stirling:2011ya}. There is some natural enhancement, 
however, because the particles carry additional spin degree of freedom 
when compared to spin-$\frac{1}{2}$ fermions. 

We find that for the 
7 TeV run of the LHC, the pair production of a colored spin--$\sp$ 
exotic fermion has cross sections in excess of a few hundred 
femtobarns (fb) for masses as high as 600 GeV. At the current run of the LHC, 
with a center of mass energy of 8 TeV, cross sections in excess 
of 100 fb are obtained for masses up to 750 GeV. Therefore a strong case 
can be made to search for such exotics in the current and 
upcoming LHC data, just as that being done for coloron like particles. 

Any search for these exotics would crucially depend on how the 
particle decays and what is produced in the final state so  
let us now discuss how these particles will decay. Higher dimension-five 
operators would lead to interactions between the massive spin--$\sp$ states and 
the spin--$\frac{1}{2}$ states such 
as \cite{Stirling:2011ya}
\begin{equation}\label{dim5}
\mathcal{L}_{dim-5} = i\frac{g_s}{\Lambda}\bar{\psi}_\alpha\left( 
               g^{\alpha\beta} + A \gamma^\alpha \gamma^\beta \right) 
           \gamma^\nu T^a \frac{(1\pm \gamma_5)}{2} \xi F^a_{\beta\nu} + H.C.
\end{equation}
where $F^a_{\beta\nu}$ represents the field tensor of the gauge field and 
$\xi$ is the spin--$\frac{1}{2}$ fermion.  $\Lambda$ determines the 
scale of some new physics  which, for example, could be the scale which 
remedies the unitarity violation seen in the cross section. Note that large 
values of scale $\Lambda$ would imply that the interaction strength weakens. 
We will assume that the colored 
spin--$\sp$ will decay promptly to a gluon and a spin--$\frac{1}{2}$ 
fermion (which in our case is a SM quark) with 100\% branching probability. 
So there is no need for us to calculate a branching ratio and thus no need to
use Eq.(\ref{dim5}).  The only thing we need is for $\Lambda$ to be large enough such 
that Eq.(\ref{dim5}) does not change the production cross section significantly.

Thus if the quantum numbers dictate a decay to a particular family of quarks we 
can have three different scenarios corresponding to the decay of the 
spin--$\sp$ particle to one of the three SM quark families, 
a light SM quark and a gluon ($Q_{3/2}\to q g$),
or a heavy quark and a gluon ($Q_{3/2}\to b g$ 
or $Q_{3/2}\to t g$).  We will now analyze each of these signals and the corresponding 
SM background representative of the type of decay.

\section{\label{sec:signals} Signals at the LHC}
\subsection{\label{sec:4j}Four jet final state}
As mentioned above we assume that the spin--$\sp$ colored fermion can decay to a SM quark 
and a gluon. If the quark happens to belong to the first two families of the 
SM quarks, then these quarks will hadronize and form jets, as will the gluons, 
leading to four jets in the final state. All the 
jets will carry large transverse momenta ($p_T$) as they are byproducts of a 
heavy particle decay. However, with final states only comprised of jets the signal
will be overwhelmed by the huge QCD background which would also be 
characterized  by high $p_T$ jets. Therefore to extract the signal from the huge 
\begin{table}[b!]
\begin{tabular}{c|c|c|c|c|c|c|c}
\cline{2-7} & \multicolumn{6}{c|}{Signal cross-section (fb) } & \\
\cline{2-7} 
 &  \multicolumn{6}{c|}{$M$ (GeV) } &  \\ \cline{1-1} \cline{8-8}
\multicolumn{1}{|c|} {$ \bm{p_T}$ cut (GeV) } 
& 500  & 600 & 700 & 800 & 900 & 1000 
& \multicolumn{1}{c|}{SM background (fb)}  \\ \hline 
& \multicolumn{6}{c|}{$\bm{\sqrt{s}=7}~\texttt{TeV}$ } &  \\ \cline{1-8} \hline
\multicolumn{1}{|c|} {200} 
& 326. & 124. & 48.6 & 18.8 & 7.2 & 2.8 &  \multicolumn{1}{c|}{11900.} \\ \hline
\multicolumn{1}{|c|} {250} 
&  134.  & 51.9 & 24.9 &   11.5 & 5.1 & 2.1 &  \multicolumn{1}{c|} {2420.} \\ \hline
\multicolumn{1}{|c|} {300} 
&   65.2 & 21.0 & 10.1 &   5.7 & 3.0 & 1.5 &  \multicolumn{1}{c|} {577.} \\ \hline
& \multicolumn{6}{c|}{$\bm{\sqrt{s}=8}~\texttt{TeV}$ } &  \\ \cline{1-8} \hline
\multicolumn{1}{|c|} {300} 
&   194. & 61.2 & 27.6 &  15.1 & 8.1 & 4.1 &  \multicolumn{1}{c|} {1270.} 
\\ \hline
\multicolumn{1}{|c|} {350} 
& 106. & 32.2 & 12.6 &   6.6 &  4.1 & 2.4 &  \multicolumn{1}{c|} {377.} 
\\ \hline
\multicolumn{1}{|c|} {400} 
& 58.1 & 17.6 & 6.5 &   3.0 &  1.8 & 1.2 &  \multicolumn{1}{c|} {118.} 
\\ \hline
& \multicolumn{6}{c|}{$\bm{\sqrt{s}=14}~\texttt{TeV}$ } &  \\ \cline{1-8} \hline
\multicolumn{1}{|c|} {400} 
& 4842. & 1549. & 569.4 & 242.2 & 120.8 & 69.7 &  \multicolumn{1}{c|}{3013.} \\ \hline
\multicolumn{1}{|c|} {450} 
& 3271. & 1074. & 399.7 & 167.6 & 79.5 & 43.3 &  \multicolumn{1}{c|} {1315.} \\ \hline
\multicolumn{1}{|c|} {500} 
& 2184.3 & 746.9 & 280.8 &  117.6 & 54.9 & 28.4 &  \multicolumn{1}{c|} {609.2} \\ \hline
\end{tabular}
\caption{\textit{The signal cross section for the $4j$ final state coming from 
the pair production of spin--$\sp$ quarks of mass M with 
$\sqrt{s}=7, 8~\text{and}~14$ TeV 
as the cut on the transverse momenta of the jets is varied. Also shown 
is the QCD background which has been estimated 
using {\texttt Madgraph 5} \cite{mad5}.}} 
\label{tab:4j}
\end{table}
background, one needs to devise some specific conditions on the kinematics of 
the final state particles and also put the focus on to the uniqueness of the signal 
coming from the new particles. The most obvious feature that the signal will exhibit 
is a peak in the invariant mass distribution of a pair of jets coming from the decay 
of the spin--$\sp$ quark. In comparison, the QCD background would trail off for 
high invariant mass values of the dijet. This signal 
could be mimicked by other new physics scenarios where new colored particles 
produced in pairs decay hadronically to dijets.  In fact the CMS Collaboration has 
made an initial analysis on such particles at LHC with $\sqrt{s}=7$ TeV using 
2.2 fb$^{-1}$ of integrated luminosity and put a lower limit on the 
mass of coloron-type particles to be 580 GeV \cite{cmsnote}.  We have used the 
CMS analysis to obtain an effective lower limit of $M \sim 490$ GeV on the mass of a spin--$\sp$ quark 
which decays into a light quark and gluon.  Note 
that the bound is lower than the coloron mass bound because for similar 
masses the pair production cross section for spin--$\sp$ quarks is smaller than 
the pair production of colorons. 

The search strategies at CMS did not include stronger cuts on 
the $p_T$ of the jets, which should further suppress the large QCD background  
for the $4j$ final state. In Table \ref{tab:4j} we summarize the signal cross section 
with different set of $p_T$ cuts on the jets in the final state and also highlight how 
the cuts affect the QCD background. In addition to the $p_T$ cut, the jets must lie 
within the rapidity gap of $|\eta_j|<2.5$ and the jets are isolated in the $(\eta,\phi)$ 
plane satisfying $\Delta R_{jj}> 0.5$, where $\Delta R$ is defined as 
$\Delta R=\sqrt{(\Delta\eta)^2+(\Delta\phi)^2}$. A minimum cut on the invariant mass
of each dijet pair has been also implemented for both signal and background, given 
by $M_{jj}>10$ GeV. As one would expect, for stronger requirements on the jet 
$p_T$, the QCD background begins to fall off rapidly. The signal is affected 
more by the $p_T$ cuts for smaller values of the spin--$\sp$ quark mass
because the jets have higher $p_T$ if they come from the decay of 
heavier  spin--$\sp$ quark. The numbers in 
Table \ref{tab:4j} demonstrate this, as stronger cuts are shown to effect the 
background more by suppressing it at times by more than 90\% which improves the 
signal to background ratio significantly. Thus with 2.2 fb$^{-1}$ integrated luminosity(L)
at $\sqrt{s}=7$ TeV, and a $p_T$ cut of 200 GeV, 
we find that the 
ratio of signal to square root of background, $S/\sqrt{B}\equiv\,L\sigma_s/\sqrt{L\sigma_b}$ is about $4.4$ for a spin--$\sp$ quark with mass 
$M=500$ GeV which suggests a significant improvement in the mass reach for 
such exotic particles. At $\sqrt{s}=7$ TeV, the stronger cuts are not helpful 
as they also suppress the signal by a large amount. 

It is worth pointing out here that our 
analysis, done at the 
leading-order parton level, does not correspond to the exact numbers seen at the 
experiments as no detector level effects have been included. However, after accounting 
for the suppression in events due to hadronization and fragmentation effects, detector 
efficiencies and acceptance, the strong cuts would still help in improving the mass reach 
for colored particles which are pair produced and decay hadronically to a pair of jets.   
      
\subsection{\label{sec:bbjj}Final state with two $b$--jets and two light jets} 
In this section we consider the scenario where the spin--$\sp$ quark quantum numbers 
dictate its decay to a bottom quark and a gluon so the pair produced  spin--$\sp$ 
quarks lead to a final state with two $b$-jets and two light jets ($2b2j$) all carrying 
large transverse momenta. This final state is already included in the $4j$ analysis when 
no heavy flavor tagging is applied on the events. However, recent analysis at both ATLAS
and CMS have shown that a very high efficiency for $b$-tagging may be obtained 
\cite{atlasbtag,cmsbtag}. Dependent on the transverse momenta of the b-jets, the 
efficiencies could be as high as 70\% for jets with $p_T> 100$ GeV. So even though we lose
part of the events due to limited efficiencies, the QCD background is reduced significantly
as the $b$-jet production forms a small subset of the full $4j$ background. On the other
hand, the pair production cross section for the spin--$\sp$ quark remains unaffected even if 
its quantum numbers correspond to a bottom quark. Therefore, the signal events will benefit 
from such flavor tagging and improve the signal to background ratio.

\begin{figure}[h!]
\centering
\includegraphics[width=2.1in]{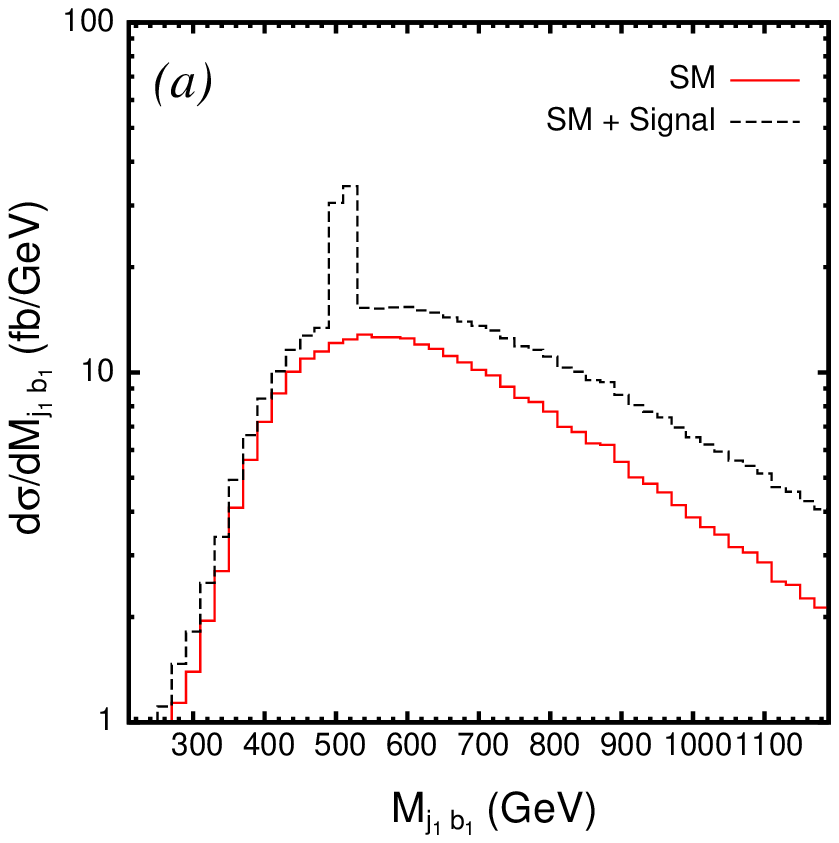}
\includegraphics[width=2.1in]{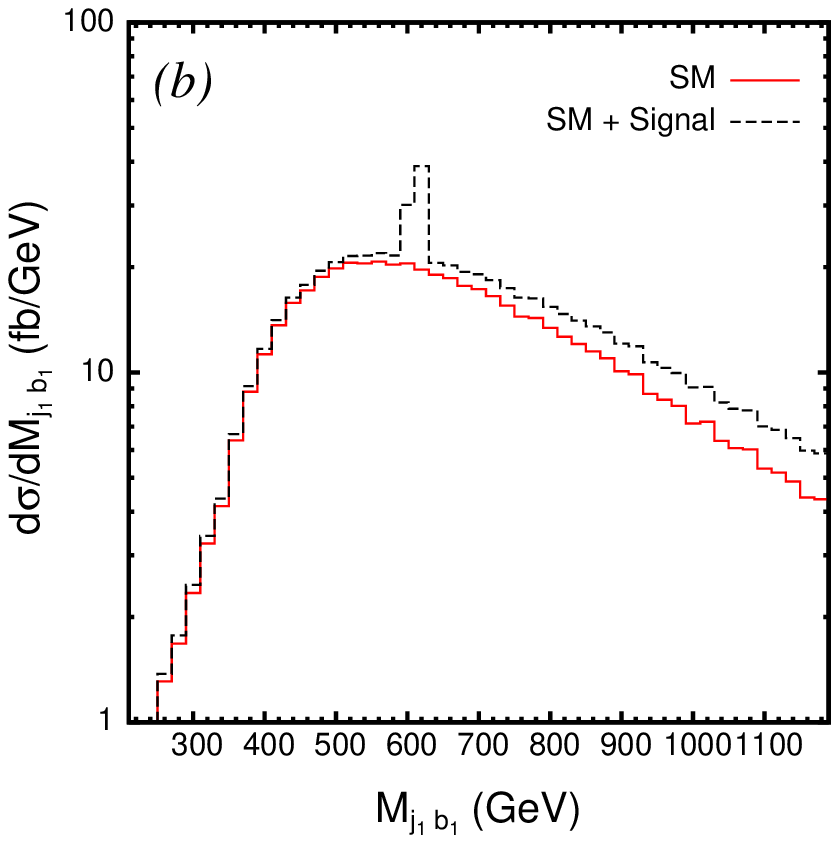}
\includegraphics[width=2.1in]{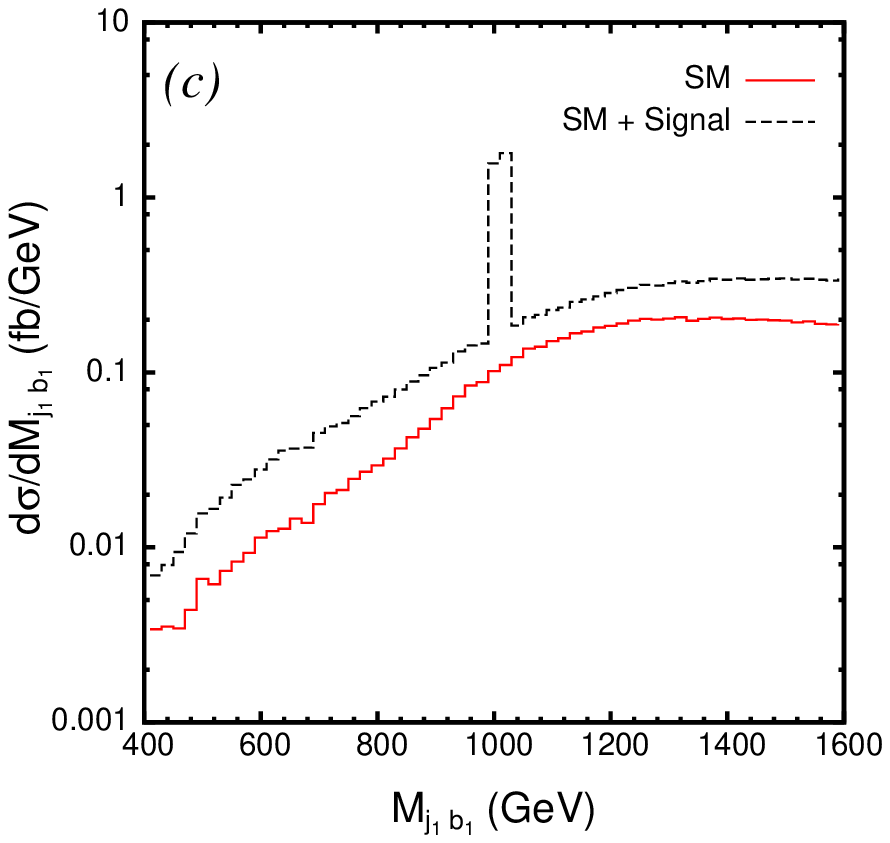}
\caption{\textit{The invariant mass distribution of the leading jet and leading $b$-jet for 
the SM background and the superposed signal coming from the production of spin--$\sp$ quarks 
with the SM background. Distributions are shown for three different values of mass of the  
spin--$\sp$ quark and at different center of mass energies, viz. (a) $M=500$ GeV; $\sqrt{s}=7$ TeV,
(b) $M=600$ GeV; $\sqrt{s}=8$ TeV and (c) $M=1$ TeV; $\sqrt{s}=14$ TeV.} }
\label{fig:MinvJ1B1}
\end{figure}
\begin{figure}[h]
\centering
\includegraphics[width=2.1in]{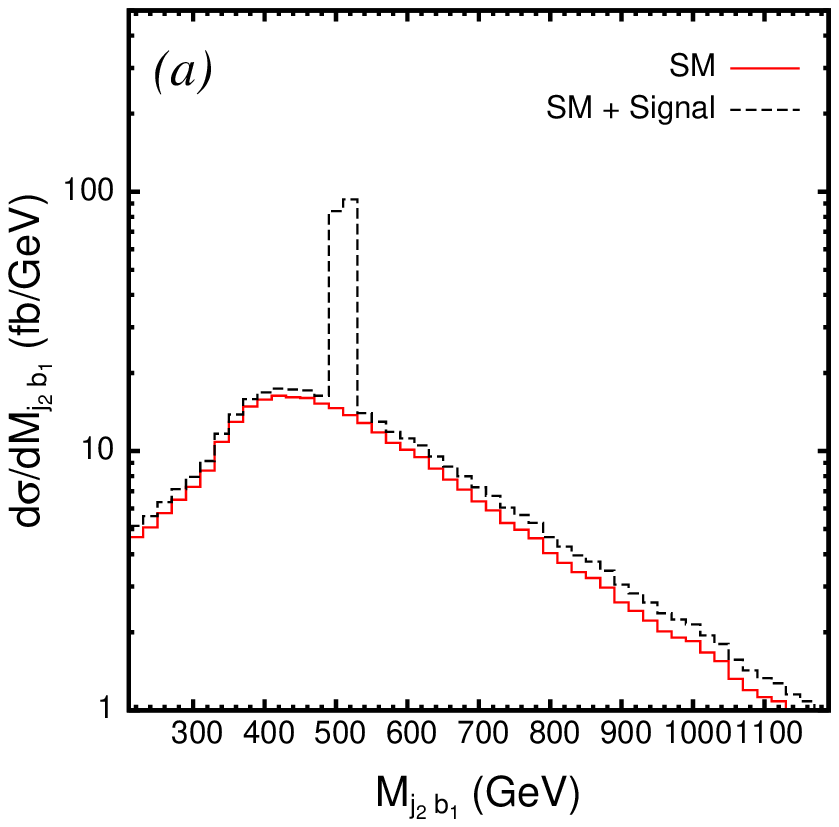}
\includegraphics[width=2.1in]{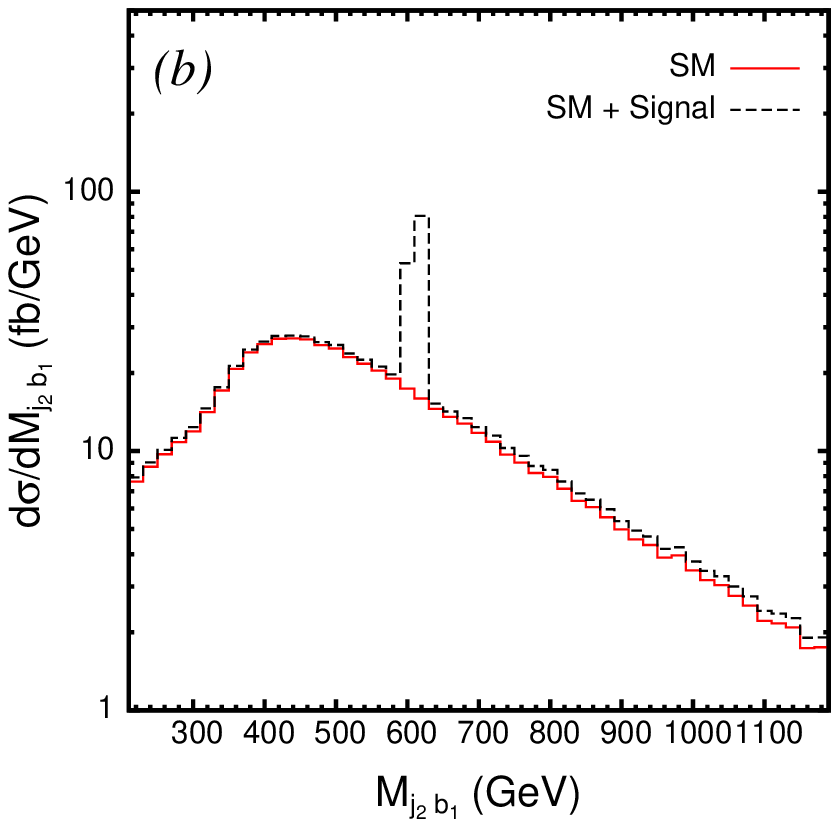}
\includegraphics[width=2.1in]{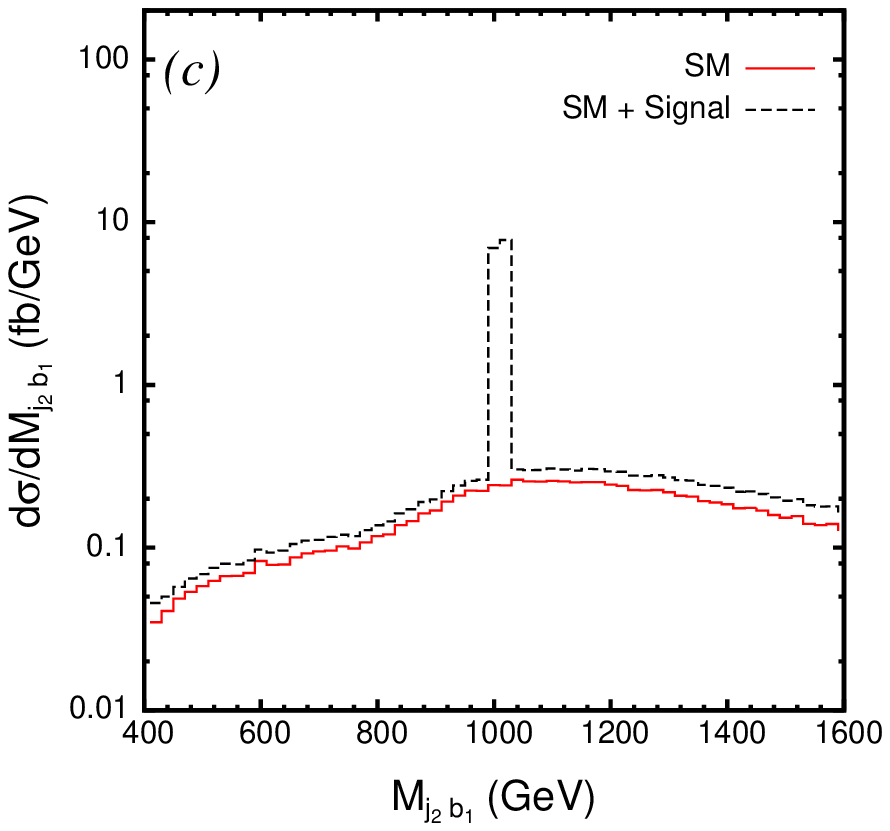}
\caption{\textit{The invariant mass distribution of the sub-leading jet and leading 
$b$-jet for the SM background and the superposed signal coming from the production of 
spin--$\sp$ quarks with the SM background. The choices of $M$ and $\sqrt{s}$ are the same as in 
Fig.\ref{fig:MinvJ1B1}.} }
\label{fig:MinvJ2B1}
\end{figure}

In Fig.\ref{fig:MinvJ1B1} and Fig.\ref{fig:MinvJ2B1} we plot the invariant mass distribution 
of the two final state jets with the leading $b$-jet. Note that in the analysis for a resonant particle,
the resonance is not seen in the two $b$-jet invariant mass but is seen in the light quark jet and 
$b$-jet. This will reduce the QCD background significantly. We plot the invariant mass distribution
for three different values of the spin--$\sp$ quark mass and at three different center of mass 
energies. Both the light quark jets and the $b$-jets are ordered 
according to their $p_T$ and we call the leading $b$-jet as $b_1$ and the subleading $b$-jet 
as $b_2$ with similar notation for the light quark jets. The events used in the plots
presented in Fig.\ref{fig:MinvJ1B1} and Fig.\ref{fig:MinvJ2B1} for both the signal and the 
background satisfy the following kinematic selection cuts:
\begin{itemize}
\item  Both the light quark jets and $b$-jets have a minimum transverse momenta 
$p_T>150$ GeV and lie within the rapidity gap of $|\eta|<2.5$.
\item To resolve the final states in the detector they should be well separated. To
achieve this we require that they satisfy $\Delta R_{ij} > 0.7$ with $i,j$ representing 
the $b$-jets and the light quark jets. As above the variable $\Delta R_{ij}$ defines the 
separation of two particles in the ($\eta,\phi$) plane of the detector with  
$\Delta R_{ij}=\sqrt{(\eta_i-\eta_j)^2+(\phi_i-\phi_j)^2}$, where 
$\eta$ and $\phi$ represent the pseudo-rapidity and azimuthal angle 
of the particles respectively. 
\item To suppress large contributions of gluon splitting into two ($b$) jets we 
demand that the minimum invariant mass of two ($b$)-jets satisfy 
$M^{inv}_{ij} > 10$  GeV. 
\item We also demand that there are no additional jets with $p_T > 150$ GeV.
\end{itemize}

A clear resonance is observed in both the $M_{j_1b_1}$ and $M_{j_2b_1}$ distributions 
in the bin corresponding to the spin--$\sp$ quark mass. It is interesting to observe that 
both the leading and subleading jet forms a resonance in the invariant mass with the leading
$b$-jet. As we have ordered the jets according to their $p_T$, their respective points of 
origin become immaterial and therefore both the combinations show an invariant mass peak.
However, the subleading jet gives the more pronounced peak with the leading $b$-jet which
seems to make it the favorable combination.      

We have used three different values for the spin--$\sp$ quark mass, 
$M=500$ GeV, $600$ GeV, and $1$ TeV at $\sqrt{s}=7,8$ and $14$ TeV respectively. 
As the larger center of mass energy gives a bigger pair production cross section
(Fig.\ref{prodfig}), we choose larger values for the spin--$\sp$ quark mass for higher
$\sqrt{s}$ to show that the signal will be significantly greater even for the larger values of mass
which are inaccessible with lower center of mass energies. We use the same
set of kinematic cuts for the analysis done at $\sqrt{s}=7$
and $8$ TeV. However, as 
in the case of $4j$ final states, stronger cuts on the 
transverse momenta of both the $b$-jet and  the light quark jet would be useful in improving 
the signal to background ratio. We therefore modify the cut on transverse momenta and 
demand that $p_T>400$ GeV for the jets at $\sqrt{s}=14$ TeV.  
For our analysis of both the signal and background, we have considered 
a $b$-tagging efficiency of 50\% while the mistag rate for light quark jets tagged as 
$b$-jets is taken as 1\%. Both the $b$-tag efficiency and the 
mistag rates are dependent on the transverse momenta ($p_T$) and 
rapidity ($\eta$) and our choices do not include these effects. To do 
such detailed analysis one would also need to include various other systematics 
including showering and hadronization effects at the LHC and detector-level 
simulations which is beyond the scope of this work. So we assume that our choice 
for the efficiencies and the mistag rate is a good approximation when averaged over the 
entire range of transverse momenta for the quarks within the allowed rapidity gap.
 
With the above set of cuts
the  signal cross section for different values of the spin--$\sp$ 
mass along with the SM background are shown in Table \ref{tab:2b2j}. 
When compared with $4j$ analysis, the reach for spin--$\sp$ quarks in the $2b2j$ channel is
found to be improved significantly. For example, for $M=1$ TeV with an integrated luminosity 
of 10 fb$^{-1}$ for $\sqrt{s}=14$ TeV and a $p_T> 400$ GeV cut 
on the jets, the $S/\sqrt{B}\simeq 4$ in the $4j$ final state while 
it becomes $S/\sqrt{B}\simeq 15$ in the $2b2j$ final state.   

\begin{table}[h!]
\begin{tabular}{|c|c|c|c|c|c|c|c|}
\cline{1-8} & \multicolumn{6}{c|}{Signal cross-section (fb) } & \\
\cline{2-7} 
$pp\to 2b2j$ &  \multicolumn{6}{c|}{$M$ (GeV) } &   SM background  (fb) \\ 
                                & 500     & 600   & 700   & 800 & 900 & 1000 &             \\  \hline
$\sqrt{s}=7~TeV$  &  182.5 &   55.0 & 17.6 &   5.9 &  2.1 &  0.7   & 351.3  \\ \hline 
$\sqrt{s}=8~TeV$  &  403.0 &  124.8& 41.6 & 14.7 &  5.5 &  2.1   & 608.9  \\ \hline 
$\sqrt{s}=14~TeV$&  584.8 & 275.4 &123.4& 57.6 & 29.7& 17.1  & 12.9  \\ 
\hline
\end{tabular}
\caption{\textit{The signal cross section for the $2b2j$ final state  
at LHC with $\sqrt{s}=7,8~\text{and}~14$ TeV for different choices of the mass 
$M$. Note that the $p_T$ cut on the jets is 150 GeV for $\sqrt{s}=7~\text{and}~8$ TeV
while it is 400 GeV for $\sqrt{s}=14$ TeV. We have included a $b$-tag efficiency $\epsilon_b=0.5$
in cross sections.}} 
\label{tab:2b2j}
\end{table}


\subsection{\label{sec:ttjj}Final state with $t\bar{t}$ and two light jets} 
Finally we specialize to the case where the spin--$\sp$ quark carries quantum numbers similar to
the top quark and therefore decays to a top quark and a gluon. This would lead 
to a $t\bar{t}$ final state with two additional jets with large transverse momenta through the 
process chain given by $pp\,\longrightarrow\,Q_{3/2}\bar{Q}_{3/2}\,\longrightarrow\,t\bar{t}gg$.


This would be a very nice signal which would not only provide a strong hint for physics beyond 
the SM but would also effect the inclusive top quark pair production if the additional 
jets are not triggered upon. However, as the production cross section of the heavier $Q_{3/2}$
particles are small compared to the pair production of $t\bar{t}$ 
(about 10\% of $\sigma_{t\bar{t}}$ for $M=400$ GeV) the new physics signal is more 
\begin{table}[h!]
\begin{tabular}{|c|c|c|c|c|}
\cline{1-5} & \multicolumn{3}{c|}{Signal cross-section  } & \\
\cline{2-4} 
$pp\to t\bar{t}jj$ &  \multicolumn{3}{c|}{$M$ (GeV) } &   SM background \\ 
($p_T^j > 100$ GeV)  & 500 & 800 & 1000 &  \\  \hline
$\sqrt{s}=7~TeV$ & 1.11 pb & 21.7 fb & 2.4 fb & 2.12 pb \\ \hline
$\sqrt{s}=8~TeV$ & 2.38 pb & 53.4 fb & 6.8 fb & 3.55 pb \\ \hline 
$\sqrt{s}=14~TeV$& 49.4 pb & 1.46 pb & 249. fb & 24.7 pb \\ 
\hline
\end{tabular}
\caption{\textit{The signal cross section for the $t\bar{t}jj$ final state coming from 
the pair production of spin--$\sp$ quarks with 
$\sqrt{s}=7, 8~\text{and}~14$ TeV for different choices of the mass 
$M$ for a fixed cut of $100$ GeV on the transverse momenta of the jets. Also shown 
is the dominant QCD background in SM which has been estimated 
using {\texttt Madgraph 5}.}} 
\label{tab:ttjj}
\end{table}
pronounced when the additional jets with high $p_T$ are triggered on. We look at the 
$t\bar{t}jj$ signal and SM background and consider a 100 GeV cut on the transverse 
momenta of the additional (nontop) jets. Note that by demanding two jets with $p_T>100$ GeV  
along with a $t\bar{t}$ pair would completely eliminate the large background coming from
the pair production of $pp\to t\bar{t}$. We generate the SM background using 
\texttt{MadGraph 5} for $pp\to t\bar{t}jj$ and  $pp\to t\bar{t}jj(+j)$ with 
some additional basic acceptance cuts of $|\eta_j|<2.5$ and $\Delta R_{jj}>0.5$.
We list the cross section for the signal and background for different 
values of the spin--$\sp$ quark mass in Table \ref{tab:ttjj}.

A quick comparison of the signal with the background shows that the although the 
background is quite large when compared to the signal for $M=1$ TeV, our experience
from the previous analysis of $4j$ and $2b2j$ signal implies that  
stronger cuts on the transverse momenta of the jets will suppress the background further. 
As before the signal will not change much for large values of $M$. 
\begin{table}[!b]
\begin{center}
\begin{tabular}{|c|c|c|}
\hline {\bf Variable} &Cut ${\mathcal C}_1$ & Cut ${\mathcal C}_2$ \\ 
\hline $p_T^{\ell,b}$ & $>10,20$ GeV & $>10,20$ GeV \\ 
\hline $\bm{p_T^j}$ & $>50$ GeV & $>200$ GeV \\ 
\hline $|\eta|$ & $<2.5$  & $<2.5$ \\ 
\hline $\bm{\Delta R_{jj}}$ & $>0.4$ & $>0.7$ \\
\hline $\Delta R_{\ell\ell,\ell j, \ell b, bj}$ & $>0.2$ & $>0.2$ \\
\hline
\end{tabular}
\caption{\textit{Two different set of cuts ${\mathcal C}_1$ and ${\mathcal C}_2$, imposed on the 
final state $\ell^+\ell^-bbjj\slashed{E}_T$ where the cuts are different only on the kinematic 
variables shown in bold. Not listed is a b-tagging efficiency of $\epsilon_b\,=\,0.5$ for both sets.}} \label{tab:cuts}
\end{center}
\end{table}

To put this in perspective let us now consider the full decay of the top quarks in the final 
state and look more closely at the signal and SM background for two different set of cuts 
on the transverse momenta of the jets. To analyze the signal we focus on the semileptonic 
decay mode of the produced top quark leading to the following final state:
\begin{align}
 p p \longrightarrow &  (Q_{3/2} \to t g) \longrightarrow (t \to b W^+)  g \longrightarrow (W^+ \to \ell^+ \nu_\ell) b g  \nonumber \\
  \hookrightarrow & (\bar{Q}_{3/2} \to \bar{t} g) \longrightarrow (\bar{t} \to \bar{b} W^-)   g \longrightarrow (W^- \to \ell^- \bar{\nu}_\ell)  \bar{b} g \nonumber \\
 \hookrightarrow \ell^+ & \ell^-  b \bar{b}  j j \slashed{E}_T 
\end{align}
where we restrict ourselves to the choice of $\ell = e, \mu$ for the charged lepton. As
it is very difficult to differentiate between $b$ and $\bar{b}$ even with heavy flavor
tagging of the jets, we are looking at a final state with a pair of 
charged leptons ($\ell^+_i\ell^-_j$),  two hard $b$-jets, two hard light quark jets and 
missing transverse momenta. We define two set of cuts which we list in Table \ref{tab:cuts}.
The results in Table \ref{tab:2l2b2j} show that going from the cuts $\mathcal{C}_1$ to the 
cuts $\mathcal{C}_2$ drastically reduces the background without much change in the signal.

The conclusions to be drawn from Table \ref{tab:2l2b2j} are the following: 
1) at $\sqrt{s}\,=\,8$ TeV the $p_T$ cuts extend the reach above $M\,=\,500$ GeV but 
well below $M\,=\,800$ TeV the signal cross-section becomes too small to be 
observed (independent of the cuts). 2) At $\sqrt{s}\,=\,14$ TeV the 
stronger $p_T$ cuts 
seem unnecessary for $M$ near $500$ GeV but become essential for $M$ equal $800$ GeV. 
At $M\,=\,1000$ GeV the cross-section is small but could be seen when the integrated luminosity 
exceeds about $200$ fb$^{-1}$.
\begin{table}[h!]
\begin{tabular}{|c|c|c|c|c|}
\cline{1-5} & \multicolumn{3}{c|}{Signal cross-section (fb) } & \\
\cline{2-4} 
$pp\to \ell^+\ell^-bbjj\slashed{E_T}$ &  \multicolumn{3}{c|}{$M$ (GeV) } &   SM background  (fb) \\ 
                             & 500 & 800 & 1000 &  \\  \hline
$\sqrt{s}=8~TeV$ & 20.1 (7.8) & 0.4 (0.3) & 0.055 (0.045) & 93.2 (2.9) \\ \hline 
$\sqrt{s}=14~TeV$& 385.9 (186.1)  & 11.2 (8.2) & 1.9 (1.6) & 522.8 (26.7) \\ 
\hline
\end{tabular}
\caption{The signal cross section for the $ \ell^+\ell^-bbjj\slashed{E}_T$ final state  
with  $\sqrt{s}=8~\text{and}~14$ TeV for different choices of the mass 
$M$ for the cuts ${\mathcal C}_1 ({\mathcal C}_2)$  shown in Table \ref{tab:cuts}. } 
\label{tab:2l2b2j}
\end{table}

\section{\label{sec:summary} Discussion and Summary}
In this work we have focused on the signals for colored spin--$\sp$ fermions at the LHC. These 
particles will have large production cross sections and can be discovered through resonances 
in different channels depending on their decay properties. We have presented 
complete analytic expressions for the parton-level matrix amplitudes and cross sections used 
in our calculations.

We considered three different scenarios for the higher spin fermion mixing with SM quarks which 
dictates the decay modes. We find that such an exotic fermion can decay hadronically to two light jets or
into a gluon and heavy quark flavors. This leads to three different final state topologies
$4j$, $2b2j$ and $t\bar{t}jj$. We did a detailed analysis of the three different cases and show that
a strong cut on the transverse momenta of the final state jets is very useful in suppressing the 
otherwise large QCD background for hadronic final states at LHC. We have compared our results with a CMS
study on $4j$ final states 
and extracted a lower bound of 490 GeV on the spin--$\sp$ quark mass. We further showed that
this reach can be improved by using stronger cuts on the $p_T$ of the jets;
the details are given in Table \ref{tab:4j}.
Given that only a limited amount of luminosity will be collected at $\sqrt{s}\,=\,8$ TeV the reach in $M$ for this final
state is between $600$ GeV and $700$ GeV.  At $\sqrt{s}\,=\,14$ TeV the reach easily exceeds $M\,=\,1$ TeV. 

We 
then considered the case where the spin--$\sp$ quark decays to a gluon and a bottom quark and 
showed, in Figs. \ref{fig:MinvJ1B1} and \ref{fig:MinvJ2B1}, that
the event characteristics of such a final state leads to a clear invariant mass peak when 
the $b$-jet is paired with the gluon jets. We also showed,
in Table \ref{tab:2b2j}, that the SM background is suppressed
in this final state which would lead to a better reach for spin--$\sp$ quark mass. 

Finally 
we focused on the signal where the spin--$\sp$ quark decays to a top quark and gluon where the
signal and background are shown, for nominal cuts, in Table \ref{tab:ttjj}. 
The background for a top pair with two additional radiated gluons is seen to be large. However, as shown in Table \ref{tab:2l2b2j},
this background can be greatly reduced by appropriate cuts which include a strong $p_T$ requirement on the jets. 
For $\sqrt{s}\,=\,14$ TeV these cuts extend the observation reach
to $M\,=\,1000$ GeV. 

\acknowledgments
This research was supported in part by the United States Department of Energy under grants No. DE-FG02-04ER41306
and No. DE-FG02-12ER41830.

\end{document}